\documentclass[aip,cha,preprint]{revtex4-1}
\usepackage{amsmath}
\usepackage{amssymb}
\usepackage{graphicx}
\usepackage{dcolumn}
\usepackage{bm}
\usepackage{epsfig}
\usepackage{natbib}

\newcommand{\ds}{\displaystyle}

\newcommand{\der}[2]{\displaystyle\frac{d #1}{d #2}}
\newcommand{\dder}[2]{\displaystyle\frac{d^2 #1}{d #2^2}}
\newcommand {\e} {\varepsilon}

\def\w{\omega}
\def\W{\Omega}

\newcommand{\new}[1]{\textbf{#1}}
\def\reff#1{(\ref{#1})}

\begin{document}
\title{
Collective Phase Chaos in the Dynamics of Interacting Oscillator Ensembles}

\author{Sergey P.~Kuznetsov}
\affiliation{Kotel'nikov's Institute of Radio-Engineering and Electronics of RAS, 
Saratov Branch, Zelenaya str., 38, Saratov, 410019, Russian Federation}
\affiliation{Institute of Physics and Astronomy, University of Potsdam, 
Karl-Liebknecht-Str. 24/25, 14476 Potsdam-Golm, Germany}
\author{Arkady Pikovsky}
\affiliation{Institute of Physics and Astronomy, University of Potsdam, 
Karl-Liebknecht-Str. 24/25, 14476 Potsdam-Golm, Germany}
\author{Michael Rosenblum}
\affiliation{Institute of Physics and Astronomy, University of Potsdam, 
Karl-Liebknecht-Str. 24/25, 14476 Potsdam-Golm, Germany}

\date{\today}

\begin{abstract}
We study chaotic behavior of order parameters in two coupled 
ensembles 
of self-sustained oscillators. Coupling within each of these ensembles is 
switched on and off alternately, while the mutual interaction between these 
two subsystems is arranged through quadratic nonlinear coupling. 
We show numerically that in the course of alternating 
Kuramoto transitions to synchrony and back to asynchrony, 
the exchange of excitations between
two subpopulations proceeds  in such a way that 
their collective phases are governed by an 
expanding circle map similar to the Bernoulli map. We perform the Lyapunov
analysis of the dynamics and discuss  
finite-size effects.
\end{abstract}

\pacs{
  05.45.Ac 	Low-dimensional chaos,
  05.45.Xt 	Synchronization; coupled oscillators }% PACS, the Physics and Astronomy                         % Classification Scheme.
\keywords{Oscillator populations, Kuramoto transition, hyperbolic chaos, robust chaos}%Use showkeys class option if keyword
                              %display desired
\maketitle

\begin{quotation}
Behavior of high-dimensional nonlinear systems of different nature (e.g. 
in hydrodynamics, 
electronics, neurodynamics, laser physics and nonlinear optics)
in many cases can be treated in terms of a cooperative action of 
a large number 
of relatively simple elements, such as interacting oscillators. 
Quite often the essential dynamics is 
low-dimensional: even
for a very large number of elements (and also in the thermodynamic limit
where this number tends to infinity) one can find a few degrees of
freedom that describe the macroscopic evolution. A famous example is the
Kuramoto model where the global variable -- complex order parameter --
\new{undergoes} a Hopf bifurcation corresponding to self-synchronization in the
ensemble of oscillators. Here we report on a modification of the
Kuramoto model where the behavior of the global variables becomes
chaotic. Our model consists of two populations that \new{undergo} Kuramoto-type
transitions, where due to an external modulation of the coupling
strength the oscillators synchronize and desynchronize 
alternately in both subpopulations.
The operation of the whole system 
consists of alternating patches of activity of the subensembles, in the sense 
that their order parameters (complex mean fields) vary between small and
large magnitudes.
Due to a specially constructed additional nonlinear coupling between the
subpopulations,  the dynamics of the phases of the complex order parameters 
is described stroboscopically   
by an expanding circle map (the Bernoulli map). Apparently, such systems 
may be designed on a base of ensembles 
of electronic devices, like arrays of Josephson junctions, 
or with nonlinear optical systems, such as arrays of semiconductor lasers.
\end{quotation}

\section{Introduction}

A long-time challenging problem motivating development of  nonlinear 
science 
concerns complex behavior of systems characterized by a large number of 
degrees of freedom. Such systems are of interest in various fields, e.g. in 
hydrodynamics (the problem of turbulence), in laser physics and nonlinear 
optics, electronics, neurodynamics \cite{Cross-Hohenberg-93}. 
The functioning of such systems 
often can be thought of as cooperative action of a large number of 
relatively simple elements, such as oscillators, with different types of
interaction between 
them~\cite{Kuramoto-75,Kuramoto-Nishikawa-87,Pikovsky-Rosenblum-Kurths-01,Acebron-etal-05}.
In many cases, nevertheless, the 
dynamics can be effectively treated as low-dimensional in terms of 
appropriate collective modes 
\cite{Takeuchi-Ginelli-Chate-09,Kuramoto-75,Kuramoto-Nishikawa-87,%
Pikovsky-Rosenblum-Kurths-01,Watanabe-Strogatz-94,Ott-Antonsen-08}.
In this context, an interesting question 
arises about a possibility of implementing various known types of 
low-dimensional complex dynamic behavior on the level of the collective 
modes.

One basic textbook example of strong chaos is Bernoulli map, or expanding 
circle map \cite{Hasselblatt-Katok-03}.
In general form, the map is $\varphi _{n + 1} = M\varphi _n 
\,(\bmod\  2\pi )$, where $M$ is an integer larger than $1$, and $\varphi $ is an 
angular variable. With initial condition for $\varphi / 2\pi $ represented 
in the numeral system of base $M$ by a random sequence of digits, the dynamics 
will correspond to a shift of this sequence by one position to the left on 
each next step of the iterations. It means that the representative point 
$\varphi_{n}$ will visit in random manner $M$ equal segments partitioning 
the circle, in accordance with the mentioned sequence of the digits.  
This is 
just what we call chaos. The sensitive dependence of the dynamics on initial 
conditions is characterized quantitatively by a positive Lyapunov exponent 
$\Lambda = \ln M$. 

In a recent series of papers \cite{Kuznetsov-05,Kuznetsov-Seleznev-06,%
Kuznetsov-Pikovsky-07}
it was shown how to implement the 
dynamics described by the Bernoulli map in low-dimensional systems of 
alternately excited 
 self-sustained oscillators.
%It is better that self-sustained oscillators because of avoiding
%the word "sustained" that gives occasion to think about 
%sustained self-oscillations, although this is not the case! 
In these systems
the existence of 
uniformly hyperbolic chaotic attractors of Smale -- Williams 
type (see, e.g., Ref. \cite{Katok-Hasselblatt-95}) was 
established~\cite{Kuznetsov-Sataev-07,Wilczak-10}.

Chaotic nature of the dynamics reveals itself in the chaotic 
evolution of the phases of oscillations generated at successive 
stages of excitation of the subsystems. The purpose of  the
present article is to 
demonstrate a possibility of chaotic behavior of similar nature at the level of 
collective variables for multidimensional systems represented by ensembles 
of oscillators.

We do not start here with some arrangement motivated by a particular
application, but construct an idealized example
that is simple and convenient for the theoretical description. 
The model is composed of two similar ensembles 
of 
%non-identical 
 self-sustained
oscillators with their natural frequencies distributed  in some range; 
within each of these ensembles the global 
coupling is switched on and off 
alternately. 
Interaction between these two subensembles is arranged with a 
special type of additional coupling through mean fields
characterized by quadratic nonlinearity. 

Due to presence of the global coupling, each ensemble can undergo the 
Kuramoto transition: as oscillators synchronize, the collective field 
emerges with notable amplitude and definite phase of its oscillations. As the 
coupling is turned off, the oscillators desynchronize because of the 
frequency detuning of individual oscillators, and the collective field 
disappears.
The operation of the whole system consists of alternating activities of two 
ensembles 
%represented by a corresponding alternating meandering
with the corresponding alternating meandering 
of their order parameters. Due to an additional coupling, 
the excitation is transmitted from one ensemble to another, and 
back, so that the phase of this collective excitation in the course of the 
process evolves in accordance with the expanding circle map mentioned
above. 

One of the goals of this paper is to compare the global realization
of hyperbolic chaos, as outlined above, with other types of collective chaos
in ensembles of dynamical systems. Postponing a detailed
discussion to the conclusion, we mention here that collective chaos has
been mainly studied in two different contexts: in ensembles of 
maps~\cite{Shibata98a,Cencini99,Pikovsky-Kurths-94a}, where individual
elements are chaotic, and in populations of
oscillators (which as individual elements are, contrary to the case of
maps, nonchaotic), either with a distribution of natural 
frequencies~\cite{Matthews-Strogatz-90,Matthews-Mirollo-Strogatz-91} or
identical~\cite{Hakim-Rappel-92,Nakagawa-Kuramoto-93,Nakagawa-Kuramoto-94,%
Nakagawa-Kuramoto-95,Takeuchi-Ginelli-Chate-09}. In this context our
study is closely related to that
in~\cite{Matthews-Strogatz-90,Matthews-Mirollo-Strogatz-91}, because we
also consider non-identical oscillators; the difference is that we
organize the coupling in a special way to ensure desired properties of
collective chaos.

\section{Basic models}
In this Section we formulate models of interacting oscillator ensembles, 
using three different levels of reduction. 
First, we consider the equations in an ``original'' form, where each oscillator is described 
by the van der Pol equation. Next, we exploit the method of averaging and formulate the model in 
terms of slowly varying complex amplitudes of oscillations. 
Finally, we proceed with 
 neglecting amplitude variations for single oscillators,
and account only for variations of the phases
on their limit cycles; that is the model of ensemble of phase oscillators.

As outlined in the Introduction, we do not study a general system of
coupled alternately synchronized ensembles, but construct a model that
should produce chaos, having specific properties (hyperbolicity).
Moreover, we want to check, how reductions to amplitude and phase
equations influence the dynamical properties. Therefore the model in
``original'' variables (Eq.~\reff{eq1}) below) looks rather
cumbersome, while its reduction to complex amplitude (Eq.~\reff{eq9}) and phase
(Eq.~\reff{eq13}) variables are rather close to the standard Kuramoto
model.

\subsection{Coupled van der Pol oscillators}

Let us consider two interacting ensembles of self-sustained oscillators. We assume that the sizes $K$ of these 
ensembles are the same, and the oscillators are described by variables $x_k$ and $y_k$, respectively, where $k=1,\ldots,K$. 
Next, we assume  that the distributions of natural
frequencies $\omega_k$ of the oscillators 
are identical for both ensembles,
with some mean frequency $\omega_0$. 
Within each ensemble, oscillators are coupled via their mean 
fields $X$ and $Y$, defined according to 
\begin{equation}
X=\frac{1}{K}\sum_k x_k\;,\qquad Y=\frac{1}{K}\sum_k y_k\;.
\label{eq-mf1}
\end{equation} 
To account for the dissipative nature of the coupling (i.e. a
tendency to 
equalize the instant states of the interacting subsystems), we assume that
it is introduced by terms in the equations containing time derivatives
of the fields $X$ and $Y$. We suppose that the coupling
strength varies in time periodically, with a slow 
period $T \gg 2\pi / \w _0 $, between 
zero and some maximum value, which exceeds the synchronization threshold
of the Kuramoto transition. 
Thus, each ensemble goes periodically through the stages of synchrony (when the coupling is large) and asynchrony 
(when the coupling is small). 
The coupling is organized in such way that these stages 
in the two ensembles occur alternately. 
Finally, there is a \textit{nonlinear} interaction between 
ensembles via the second-order mean fields 
\begin{equation}
X_2=\frac{1}{K}\sum_k x_k^2\;, \qquad Y_2=\frac{1}{K}\sum_k y_k^2\;.
\label{eq-mf2}
\end{equation}
Being  represented as sums of the squares of the original variables, 
these fields  $X_2$ and $Y_2$ contain 
components 
with the double frequency in comparison to the frequency of the 
variables $X$ and $Y$.
To ensure 
an efficient \textit{resonant interaction} we need the components with the main
oscillator frequency; to this end the coupling terms are chosen as
products   of time 
derivatives of
the fields $X_2,Y_2$  and of
an auxiliary signal $\sin\omega_0 t$.  These products then contain the
components with the basic frequency $\omega_0$.
%Because the frequency of these mean fields is twice as large as frequency of the mean fields $X,Y$, 
%we multiply $X_2$, $Y_2$ with an auxiliary signal $\sin\omega_0 t$, to ensure the efficient interaction.
The set of governing equations for the model reads: 
\begin{equation}
\label{eq1}
\begin{aligned}
\dder{x_k}{t} - (Q - x_k^2)\der{x_k}{t} + \w_k^2x_k &= 
\kappa f_1 (t)\der{X}{t} + 
\e \der{Y_2}{t}\sin \w_0 t\;, \\[2ex]
 \dder{y_k}{t} - (Q - y_k^2)\der{y_k}{t} + \w_k^2y_k &= 
 \kappa f_2 (t)\der{Y}{t} + 
\e \der{X_2}{t}\sin \w_0 t\;, 
 \end{aligned}
\end{equation}
where $k=1,2,...,K$. The functions $f_1 (t) = \cos ^2(\pi t / T)$ 
and $f_2 (t) = \sin ^2(\pi t / T)$  
describe the alternate on/off switching of the couplings
inside the ensembles. 
Parameter $Q$ determines the amplitude of each single
van der Pol oscillator; parameters $\kappa$ and $\e$ 
characterize, respectively, the
internal and mutual couplings for the ensembles.
 It is assumed that the period of coupling modulation
contains a large integer number of periods of the auxiliary signal,
i.e. $\omega_0 T/2\pi=N>>1$.

%Noteworthy, in order to ensure the dissipative character of the coupling, we implement it via time derivatives 
%of the mean fields.

\subsection{Description in terms of complex amplitudes}

The method of averaging allows us to describe weakly nonlinear 
oscillators in a simplified form, 
via the slow dynamics of the complex amplitudes;
this description is appropriate under assumptions that 
$|\w_k-\w_0|<<\w_0$, $2\pi/T<<\w_0$.
For the van der Pol oscillators (\ref{eq1}) the averaging 
can be accomplished by introducing complex 
amplitudes $a_k$, $b_k$ according to
\begin{equation}
a_k(t)=e^{-i\w_0 t}\frac{\dot{x}_k+i\w_0 x_k}{i\w_0}\;,\qquad 
b_k(t)=e^{-i\w_0 t}\frac{\dot{y}_k+i\w_0 y_k}{i\w_0}\;.
\label{eq-transf}
\end{equation}
Substituting this in Eqs.~(\ref{eq1}) and averaging over the period of fast oscillations $2\pi/\w_0$ 
(practically, this may be done simply by dropping all term in the r.h.s., 
which contain fast time dependencies 
like $\sim e^{\pm i\w_0 t}$, $e^{\pm 2i\w_0 t}$ etc.), we obtain
\begin{equation}
\label{eq9}
\begin{aligned}
 \dot {a}_k &= i\Omega _k a_k + \frac{1}{2}(Q - \vert a_k \vert^2)a_k 
+ \frac{1}{2}\kappa f_1 (t) A + \frac{1}{2}i\varepsilon B_2 \;, \\[2ex] 
 \dot {b}_k &= i\Omega _k b_k + \frac{1}{2}(Q - \vert b_k 
\vert ^2)\,b_k + \frac{1}{2}\kappa f_2 (t) B + \frac{1}{2}i\varepsilon A_2 \;.
\end{aligned}
\end{equation}
Here $\Omega_k=\omega_k-\w_0$ are the frequency 
differences with respect to the average one. 
The mean fields $A,B,A_2,B_2$ are defined similarly 
to (\ref{eq-mf1}) and (\ref{eq-mf2}) as
\begin{equation}
A=\frac{1}{K}\sum_k a_k\;,\qquad B=\frac{1}{K}\sum_k b_k\;,\qquad A_2=\frac{1}{K}\sum_k a_k^2\;,
\qquad B_2=\frac{1}{K}\sum_k b_k^2\;.
\label{eq-mf3}
\end{equation}

\subsection{Phase approximation}

One more step in simplification of the model is 
to completely neglect  the amplitude variations for single elements,
and to reduce the description 
to  that in terms of ensembles of phase oscillators. 
In this approximation the individual oscillators 
are assumed to have a constant 
amplitude  (that of the limit cycle of a single oscillator)
throughout the process, while the dynamics manifests itself only 
in the evolution of their phases. 
This is a widely used approximation 
in the theory of synchronization, 
see \cite{Kuramoto-75,Pikovsky-Rosenblum-Kurths-01}.

To  derive the phase equations, we 
substitute in Eq.~(\ref{eq9})
\begin{equation}
\label{eq11}
a_k(t) \approx R\,e^{i\theta _k(t) }\;, 
\qquad\qquad b_k(t) \approx R\,e^{i\phi 
_k (t)}\;,
\end{equation}
where $R$ is a constant, equal for all oscillators. 
We specify $R = \sqrt {Q + \kappa / 2}$ according to the following reason.
First, if one switches off the coupling between the ensembles 
setting $\e=0$, then  each single 
oscillator is described by the equation 
$\dot {a} = i\Omega a + \frac{1}{2}(Q -\vert a \vert ^2)a
+ \frac{1}{2}\kappa  A $. In 
absence of the synchronization  $A=0$, and the amplitude of 
the stationary 
self-sustained oscillations is determined from $\vert a \vert ^2=Q$. 
On the other hand, in the 
case of maximal synchronization one has $f=0$, $A = a$, 
and, hence, $\vert a\vert ^2 = 
Q + \kappa $. Since the dynamics consists of the alternating epochs of 
synchronization and desynchronization, it is reasonable to take the average 
value. \footnote{\new{For the parameters as in Figs.~\ref{fig:vdpfield},\ref{fig:cafield} 
below, we
observed that the amplitudes of individual oscillators (not to be mixed
up with the amplitudes of the mean fields presented in these figures!)
vary in the ranges $1.66-1.96$ at the edge of the band and $1.74-1.98$
in the middle of the band, while the value of $R$ according to formula
above yields $1.87$.}}

Then, from Eqs.~(\ref{eq9}),  we obtain equations for 
the phase dynamics: 
\begin{equation}
\label{eq13}
\begin{aligned}
 \dot {\theta }_k &= \Omega_k + \frac{1}{2}
\mbox{Im}\left[ {(\kappa f_1(t) U + i\varepsilon R V_2 )e^{ - i\theta_k }} \right]\;, \\ 
 \dot {\phi }_k &= \Omega_k + \frac{1}{2}\mbox{Im}\left[ {(\kappa f_2(t) V + i\varepsilon R U_2 )e^{ - i\phi _k }} 
\right]\;,
 \end{aligned}
\end{equation}
where the complex mean fields $U,V,U_2,V_2$ are defined according to
\begin{equation}
U=\frac{1}{K}\sum_k e^{i\theta_k}\;,\qquad
V=\frac{1}{K}\sum_k e^{i\phi_k}\;,\qquad
U_2=\frac{1}{K}\sum_k e^{2i\theta_k}\;,\qquad
V_2=\frac{1}{K}\sum_k e^{2i\phi_k}\;.
\label{eq-mf4}
\end{equation}

\subsection{Thermodynamic limit}

Above we have assumed the number of oscillators in the ensembles to be finite. 
Now let us write the equations in the ``thermodynamic limit'' $K\to\infty$. 
In the case we deal with, the oscillators differ only by their 
natural frequencies, 
then it is convenient 
to parameterize the oscillators by this continuous variable, i.e. 
by the frequency $\w$, and introduce a  
distribution over the frequencies characterized by a function
$g(\w)$.
Respectively, after transformation of the equations
to slow amplitudes, and in the phase approximation, 
we will use the index $\W=\w-\w_0$ designating the frequency difference.
Then, we characterize the
distribution by the density $\tilde g(\W)=g(\w_0+\W)$.
 
With these notations, the set of the 
phase equations (\ref{eq13}) can be rewritten as
\begin{equation}
\label{eq14}
\begin{aligned}
 \dot {\theta }_\W &= \W + \frac{1}{2}
\mbox{Im}\left[ {(\kappa f_1(t) U + i\varepsilon R V_2 )e^{ - i\theta_\W }} \right]\;, \\ 
 \dot {\phi }_\W &= \W + \frac{1}{2}\mbox{Im}\left[ {(\kappa f_2(t) V + i\varepsilon R U_2 )e^{ - i\phi_\W }} 
\right]\;,
 \end{aligned}
\end{equation}
where the mean fields are defined now via the integrals over the density
\begin{equation}
\begin{aligned}
&U=\int\,d\W\, \tilde g(\W) e^{i\theta_\W}\;,\quad
V= \int\,d\W\, \tilde g(\W) e^{i\phi_\W}\;,\\
&U_2= \int\,d\W\, \tilde g(\W) e^{2i\theta_\W}\;,\quad
V_2= \int\,d\W\, \tilde g(\W) e^{2i\phi_\W}\;.
\end{aligned}
\label{eq-mf5}
\end{equation}
In a similar way, the equations for the complex amplitudes and for 
the original van der Pol oscillators may be easily reformulated 
in the thermodynamic limit.

\section{Numerical evidence for collective chaos}
In this Section we present numerical studies of 
the models (\ref{eq1}), (\ref{eq9}), and (\ref{eq13}). 

 As is known from the theory of synchronization in populations 
of oscillators developed by Kuramoto, 
the properties of the synchronization transition are qualitatively 
the same for all unimodal smooth distributions of 
oscillators over their frequencies. 
In computations below we specify the distribution 
for the model (\ref{eq1}) as 
\begin{equation}
\label{eq4}
f(\w) = \left\{ 
\begin{array}{ll}
 \ds\frac{\pi }{2}\cos \frac{\pi(\w- \w_0-\Delta\w)}{2\Delta\w}, 
    & \qquad\w\in [\w_0-\Delta\w,\w_0+\Delta\w]\;,\\  0 
&\qquad{\rm otherwise\;,} 
\end{array} 
\right.
\end{equation}
and set $\w_0=2\pi$ and $\Delta\w=\frac{\pi}{8}$.
 
We  select the form (\ref{eq4}) because it allows us 
to choose the discrete set of frequencies for a 
finite ensemble according to a simple relation
\begin{equation}
\w _k = \w_0-\Delta\w + \frac{2\Delta\w}{\pi }\arccos \left[ {\frac{2k - 1}{K} - 1} \right]\;,\qquad k=1,2,\ldots K\;.
\label{eq-distr}
\end{equation}
Furthermore, absence of significant tails of the distribution 
(compared, e.g., to a Lorentzian one) 
simplifies numerical studies,  as we do not have  to bother about
non-resonant oscillators with too large or too small frequencies.
For the models (\ref{eq9}), and (\ref{eq13})  
analogous distributions were used obtained from (\ref{eq4}) and (\ref{eq-distr}) 
with the substitution $\W=\w-\w_0$.                                   

With this setup, we simulated dynamics of the models 
(\ref{eq1}), (\ref{eq9}), and (\ref{eq13}) in computations
at parameters $T=100$, $\kappa=1$, $\varepsilon=0.1$, and $Q=3$, for system sizes $K=1000$
and $K=10000$.
The main quantities of interest are the phases of the mean fields. 
For the ensembles of van der Pol oscillators we define them according to

\begin{equation}
\Phi_X=-\arctan \frac{\dot X}{\w_0 X}\;,\qquad \Phi_Y=-\arctan \frac{\dot Y}{\w_0 Y}
\label{eq-defphvdp}
\end{equation}
For the ensembles (\ref{eq9}) and (\ref{eq13}) we 
define the phases simply as the arguments of the complex mean fields
\begin{equation}
A=\vert A\vert e^{i\Phi_A}\;,\quad
B=\vert B\vert e^{i\Phi_B}\;,\quad
U=\vert U\vert e^{i\Phi_U}\;,\quad
V=\vert V\vert e^{i\Phi_V}\;.
\label{eq-defphcp}
\end{equation}

\begin{figure}[!htbp]
\centering
 \includegraphics[width=0.7\textwidth]{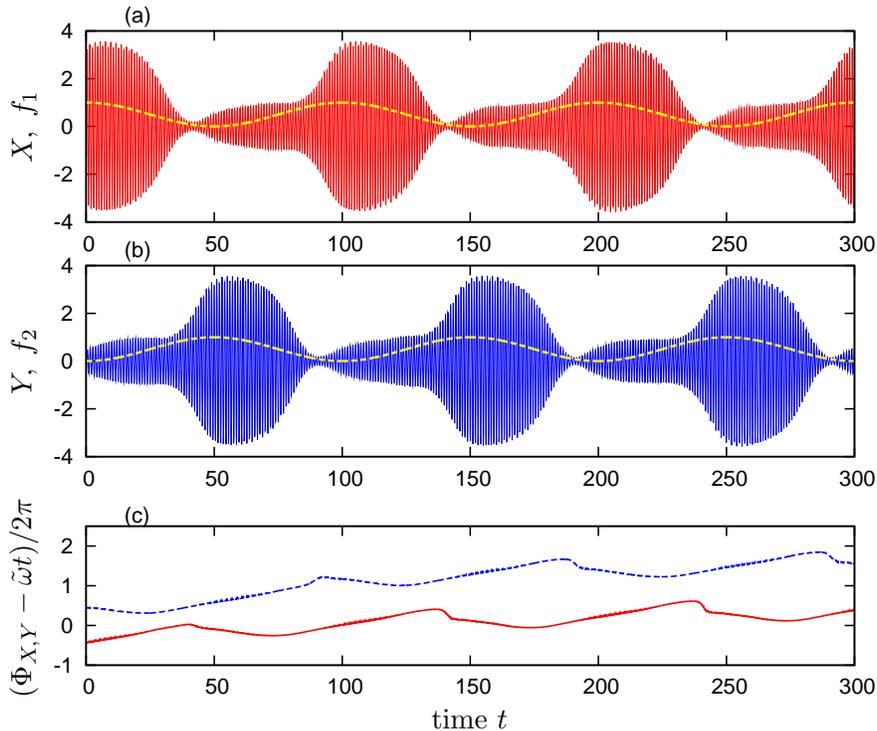}
\caption{The mean fields (panels (a) and (b)) 
and their phases (panel (c)) for two interacting 
ensembles of van der Pol oscillators (\ref{eq4}) with $K=1000$. 
The modulation function of the coupling 
$f_{1,2}$ are shown in (a,b) with dashed lines. 
 To make the picture (c) clear, we subtract a linear phase growth and plot 
the value $\Phi_{X,Y}-\tilde\w t$, where $\tilde\w=\w_0-0.2$ is chosen 
close to the empirical mean frequency $\tilde\w$ , which  
slightly differs from $\w_0$ due to the nonlinearity.
$\Phi_X-\tilde\w t$ is shown with solid red line and $\Phi_Y-\tilde\w t$ 
with dashed blue one.}
\label{fig:vdpfield}
\end{figure}

Figure~\ref{fig:vdpfield} illustrates the dynamics of the mean 
fields $X(t)$ and $Y(t)$ in the ensemble of van der Pol 
oscillators (\ref{eq1}). First, we mention that 
 because of the modulation $\sim \kappa f_{1,2}(t)$ the mean 
fields of two ensembles vary significantly: they drop nearly 
to zero in the epochs where the corresponding values of $f_{1,2}$ are small, 
and attain large values in the epochs where the coupling inside 
%the 
a population  becomes large. In panel (c) we show the 
time dependences of the phases of the mean fields $\Phi_{X,Y}$. 
One can see slight regular variations of the phases 
 correlated
%according 
to the amplitudes of the mean fields, 
and additional phase shifts 
close to
the moments of time when the corresponding mean fields nearly vanish 
(at times $t\approx 40,140,240$ for $X(t)$, 
and at times $t\approx 90,190,290$ for $Y(t)$). 
These phase shifts correspond to \textit{transfer of the phase} 
from one ensemble to another through 
their mutual coupling terms 
proportional to $\e$.

 Usually, as an ensemble of oscillators passes through the 
Kuramoto
transition from a non-synchronous to a synchronous state while the
coupling strength increases, the phase 
(potentially, of arbitrary value) 
of the arising
collective mode is determined by fluctuations
stimulating the excitation of this mode. In our setup, however, the excitation
occurs in presence of 
a small driving force
$\sim\e$ because of action of another ensemble, 
which is synchronous at that moment
and generates notable mean field.
This stimulation determines the
phase on the appearing collective mode, which accepts this
externally designated phase. This is the mechanism of the phase transfer.

Because the mutual couplings are proportional to  the 
\textit{second-order} 
mean fields 
characterized by
a doubled frequency, the phase transfer is accompanied with
doubling of the phase. To explain this, let us assume that
at the transfer of excitation from
the first to the second subensemble we have 
$X \sim \cos (\omega_0 t+\Phi)$
and, respectively, 
$\dot{X_2} \sim \sin [2(\omega_0 t+\Phi)+\mbox{const}]$.
Then, the driving force affecting the second subensemble contains
the resonance component of 
$\dot{X_2}\sin \omega_0 t \sim \cos (\omega_0 t+2\Phi+\mbox{const})$.
Respectively, the arising mean field will be of the form
$Y \sim \cos (\omega_0 t+2\Phi+\mbox{const})$.

 As doubling of the phase occurs at each transfer from one 
subsystem to another and back
through the full cycle (i.e. over the period $T$), 
as a result we expect that the phase is multiplied by 
the factor of $4$ (up to an additive constant). 

\begin{figure}[!htbp]
\centering
\includegraphics[width=0.8\textwidth]{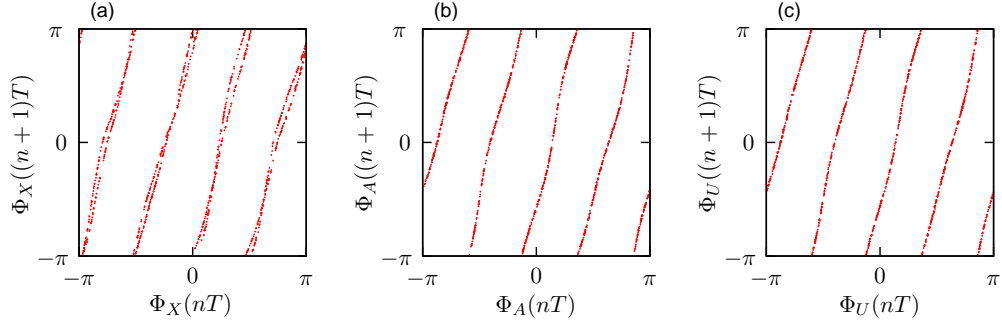}
\caption{Stroboscopic maps over the period of external modulation $T$ for:
(a) the ensembles of van der Pol oscillators (\ref{eq4}); (b) for 
the ensembles described by slow
complex amplitudes (\ref{eq9}); 
(c) for the ensembles in the phase approximation (\ref{eq13}); in all
cases $K=1000$. 
In all cases the dynamics  seems well described by the 
Bernoulli map (\ref{eq-bm}).  The observed splitting of the ``lines'' 
(the most pronounced in panel (a)) appears because of presence of transversal 
fractal structure of the attractors (see Fig.\ref{fig:pm1}):
distinct filaments of the attractor give rise do distinct filaments
on the phase iteration diagram due to imperfection
of the phase definition.
}
\label{fig:pm}
\end{figure}

To check the supposed mechanism of the phase transfer,
we constructed numerically a stroboscopic 
 iteration phase diagram
%Poincar\'e plot 
$\Phi_X(nT)\to\Phi_X((n+1)T)$, 
relating the phases at the successive moments of maximal
amplitude of the mean field $X$. 
This map, shown in Fig.~\ref{fig:pm}(a), clearly  indicates
that the transformation is indeed close to the 
Bernoulli-type map
\begin{equation} 
\Phi_X((n+1)T)=4\Phi_X(nT)+\mbox{const}\pmod{2\pi}\;.
\label{eq-bm}
\end{equation}

\begin{figure}[!htbp]
\centering
\includegraphics[width=0.7\textwidth]{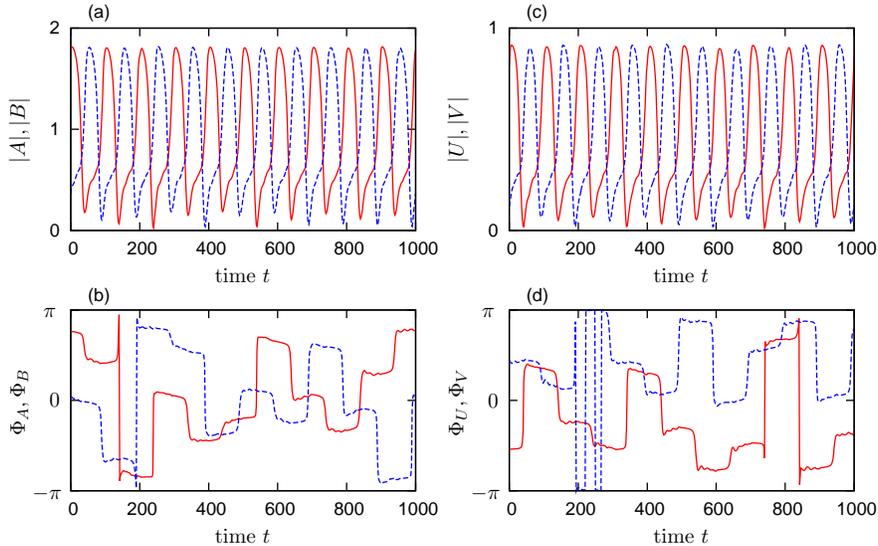}
\caption{Evolution of amplitudes and phases
for the mean fields of coupled ensembles described by 
slow complex amplitudes (panels (a,b)) and in the 
phase approximation (panels (c,d)), $K=1000$. 
Solid lines: variables $\vert A\vert, \Phi_A,\vert U\vert, \Phi_U$, dashed lines:
 variables $\vert B\vert, \Phi_B,\vert V\vert, \Phi_V$.
}
\label{fig:cafield}
\end{figure}

For the same values of the parameters we also simulated the 
 dynamics of the coupled oscillator ensembles in the slow
 complex amplitude 
 version of
%formulation 
Eq.~(\ref{eq9}) (see Fig.~\ref{fig:cafield}(a,b)), 
and in the phase approximation of Eq.~(\ref{eq13}) 
(see Fig.~\ref{fig:cafield}(c,d)). 
In these situations the 
second-order effects of amplitude-dependent frequency shifts, like those
seen in Fig.~\ref{fig:vdpfield}c, are not observed.
As a result, the phases between 
the short intervals of phase transfers are nearly constant, 
and the phase shifts at the transfers are clearly visible. 
The stroboscopic maps of the phases are shown in Fig.~\ref{fig:pm}(b,c). 
They demonstrate Bernoulli-type maps similar to that for the ensemble of van der Pol 
oscillators  of Eq.~(\ref{eq-bm}) (see Fig.~\ref{fig:pm}(a)).

%\begin{figure}[!htbp]
%\centering
%\includegraphics[width=0.5\textwidth]{../phase_ens.eps}
%\caption{
%}
%\label{fig:phfield}
%\end{figure}

\begin{figure}[!htbp]
\centering
\includegraphics[width=0.8\textwidth]{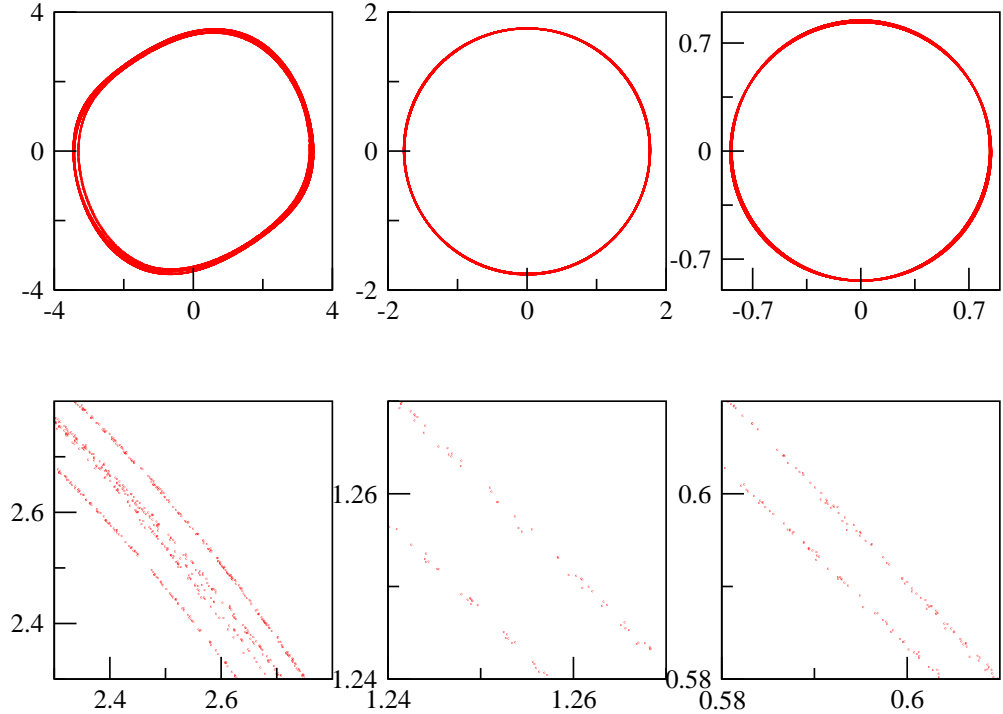}
\caption{Projections of the stroboscopic maps on the plane of the order
parameters: $(X,\dot X)$ for the van der Pol oscillators (left column);
$(\text{Re}A,\text{Im}A)$ for the ensembles described by slow
complex amplitudes (center column); $(\text{Re}U,\text{Im}U)$ for the ensembles 
in the phase approximation. Bottom row show enlargements to make the
fractal transversal structure evident. Here $K=10000$.}
\label{fig:pm1}
\end{figure}

For a further numerical characterization of the chaotic dynamics of mean
fields, we constructed the stroboscopic maps of the complex mean fields
via period of modulation $T$.  Portraits
of attractors of these maps for our three levels of
description of the ensembles are
depicted in Fig.~\ref{fig:pm1}. For a dynamical system where the phase
(or another cyclic variable)
undergoes a Bernoulli-type transformation, while in other directions in 
the phase space the
phase volume compresses,  one expects
the strange attractor to be of the Smale-Williams type, i.e. represented
by a solenoid. In a two-dimensional projection this attractor looks like
a circle with a fractal transversal structure. Fig.~\ref{fig:pm1} confirms
this picture for the dynamics of the mean field. 

\begin{figure}[!htbp]
\centering
\includegraphics[width=0.7\textwidth]{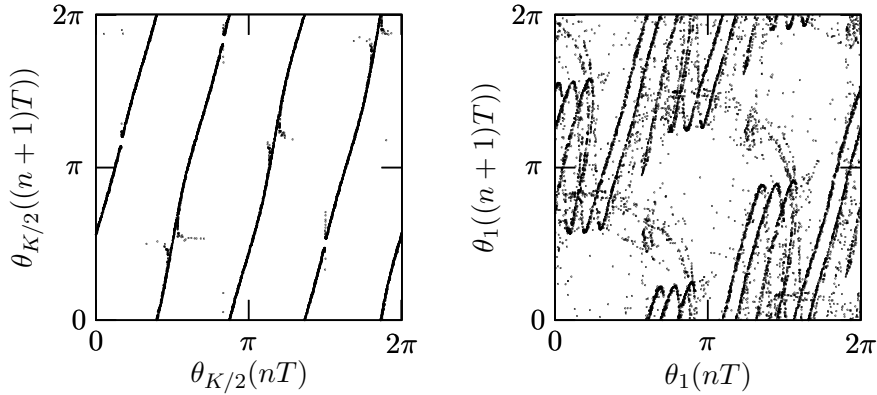}
\caption{Stroboscopic maps over the period of external modulation $T$
for individual oscillators.
Left panel: oscillator in the center of the band 
with $\omega=\omega_0$; right panel:
oscillator with $\omega=\omega_0-\Delta\omega$. }
\label{fig:pm2}
\end{figure}

We would like to stress that the Bernoulli map describes the
\textit{collective} phase (i.e. the phase of the mean field) but not
individual phases of the oscillators. Indeed,  as is evident from topological
considerations, the doubling of the phase
is only possible, if the amplitude vanishes at some stage. For the 
complex mean
field  the phase doubling is achieved by 
synchronization - desynchronization, while
 all individual oscillators always have a finite amplitude and therefore
cannot be described by the doubling Bernoulli map. We illustrate this in
Fig.~\ref{fig:pm2},  where the maps are shown similar to that
of Fig.~\ref{fig:pm}, but for the individual phases of two 
 representative oscillators
of the ensemble. Mostly close to the Bernoulli map is the behavior
of the oscillator at the center of the  frequency band, 
as this oscillator nearly
perfectly follows the mean field. Nevertheless, one can
clearly see the ``defects'' of the transformation appearing because the
``amplitude'' of the oscillator cannot vanish. The oscillator at the
edge of the band does not generally follow the phase of the mean field,
and its dynamics is far from the Bernoulli map.

\section{Finite size effects}
\begin{figure}[!htbp]
\centering
% \psfrag{xlabel}[c][c]{$K$}
% \psfrag{ylabel1}[c][c]{$\lambda_{max}$}
% \psfrag{ylabel2}[c][c]{$\lambda_{1-3}$}
% \psfrag{ylabel3}[c][c]{Period}
\includegraphics[width=0.5\textwidth]{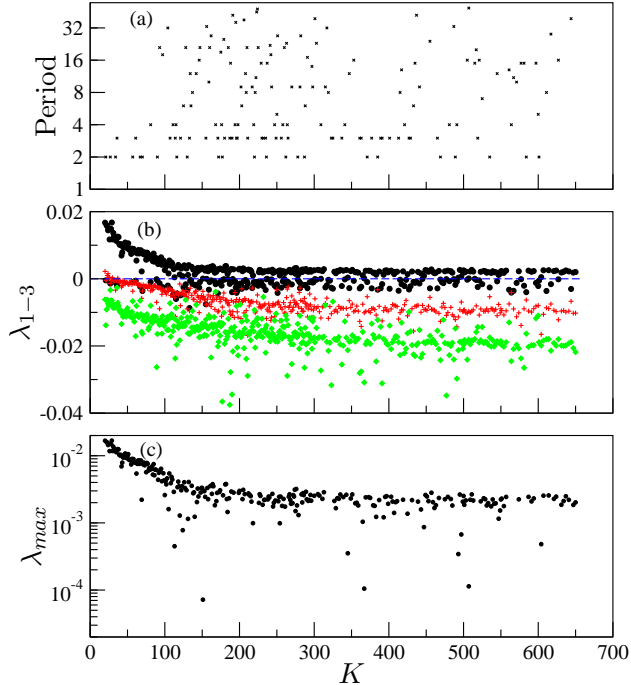}
\caption{Bifurcation diagrams for the ensembles in the complex amplitude 
formulation (\ref{eq9}) showing dynamical regimes in dependence on the
ensemble size $K$. (a): Periods for periodic regimes. The values of $K$ for
 which no period is plotted correspond to the chaotic states. (b): Three
largest Lyapunov exponents (the first: black filled circles, the second:
red pluses, the third: green diamonds). Only a few regimes with small
number of oscillators have two positive exponents. In panel (c) we show
only positive largest
Lyapunov exponents, in a logarithmic scale, to demonstrate that it does
not tend to decrease for large ensemble sizes $K$. }
\label{fig:bd}
\end{figure}

In this section we characterize collective chaos for different values of the 
ensemble size $K$. We focus here on the properties of the model
Eq.~\reff{eq9} in terms of complex amplitudes.
 First, we analyzed the stroboscopically observed
 phases of the mean field and found that for some ensemble sizes $K$ a
 periodic behavior is observed. The periods of the found periodic
 regimes are shown in Fig.~\ref{fig:bd}(a) vs $K$. The values of $K$ for
 which no period is plotted correspond to the chaotic states.
At first glance, this contradicts robustness of 
chaos expected because of its approximate description in terms
of the expanding Bernoulli map.
 Apparently, changing the ensemble size we make
an essential
perturbation
of the effective collective dynamics, so that the usual arguments of
structural stability do not apply.

\begin{figure}[!htbp]
\centering

\includegraphics[width=0.8\textwidth]{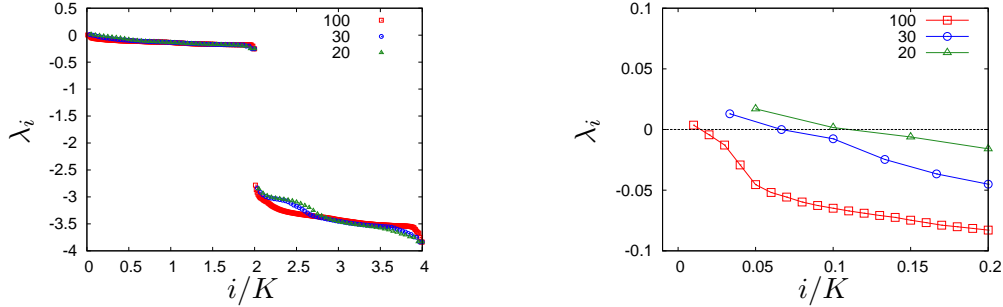}
\caption{Lyapunov exponents $\lambda_i$ plotted vs index $i/K$ for the ensembles described by 
complex amplitudes (\ref{eq9}) with $K=20,30,100$. $2K$
exponents corresponding to the phases are close to zero, while the rest
$2K$ exponents corresponding to the stable amplitudes are negative. Right panel shows the
region with small $i$, in all cases only the first exponent is positive. 
}
\label{fig:le1}
\end{figure}

The analysis via the bifurcation diagrams Fig.~\ref{fig:bd}(a) is confirmed
by calculation of the Lyapunov exponents for the ensembles. We have
performed this analysis for the ensembles described by 
complex amplitudes (\ref{eq9}) of different
 size. 
First, in
Fig.~\ref{fig:le1} we present the full spectrum of the Lyapunov
exponents for three ensemble sizes. For these parameters we observe
chaos, and in all cases only one Lyapunov exponent is positive. We
interpret this as existence of one collective chaotic mode, contrary to
the cases when many Lyapunov exponents in populations of oscillators are
positive (cf.~\cite{Nakagawa-Kuramoto-95,Takeuchi-Ginelli-Chate-09,Topaj-Pikovsky-02}).

In Fig.~\ref{fig:bd}(b) we present the three first 
Lyapunov exponents for the same data as in  Fig.~\ref{fig:bd}(a). 
 Again, like in
Fig.~\ref{fig:bd}(a), one can see that for certain system sizes the
dynamics is regular, as the largest exponent is negative. One can notice
that the largest positive Lyapunov exponent decreases with $K$ for
$K<150$, but, as panel Fig.~\ref{fig:bd}(c) shows, saturates and 
presumably does not further decrease for
larger system sizes (contrary, e.g. to the 
reported in~\cite{Popovych-Maistrenko-Tass-05}
dependence $\lambda_{max}\sim K^{-1}$ in the standard Kuramoto model), 
although periodic windows can be observed for large
$K$ as well. (Preliminary calculations demonstrate that for large $K$ 
these windows become extremely rare.)

We suggest that the observed peculiarities are specific 
just to  finite-size effects, and that they will possibly 
disappear 
in the thermodynamics limit. Indeed, the computations show that 
they become less expressed under increase of the ensemble size;
however, as may be concluded from computations,
the convergence to the large-size behavior is
slow enough.

\section{Conclusion}

We have proposed and studied a model system, which demonstrates 
chaotic dynamics 
of the phases of the mean fields. These collective variables
are described by an expanding circle map transformation in the course 
of the transfer of collective excitation alternately between two 
synchronizing and desynchronizing groups of oscillators. We have
demonstrated this on different levels of description: for the original
system of coupled van der Pol oscillators, for the model based on the
equations for slowly
varying complex amplitude, and for the phase oscillators. While collective
chaos is observed in a wide range of parameters, it is  not completely robust
with the respect to variations of the ensemble sizes: here we observe
regularity windows. These finite size effects require further
investigations. 

An interesting property of the model studied is that in the chaotic
state it has only one positive Lyapunov exponent (except for several
regimes with a very low number of oscillators in ensemble), all others
are negative. In this
respect our model differs from the ensemble of identical oscillators studied
in~\cite{Nakagawa-Kuramoto-95,Takeuchi-Ginelli-Chate-09} where the
number of positive Lyapunov exponents in the regime of collective chaos 
was macroscopic (proportional to
the size of the ensemble). It also differs the ensemble of identical Josephson
junctions~\cite{Watanabe-Strogatz-94}, where  few Lyapunov exponents are
positive but the macroscopic majority of them vanish due to partial
integrability of the system. At the moment it is not clear, which
physical properties of the systems are responsible for this difference.
Possibly a comparison with the Lyapunov spectrum in models with distribution
of frequencies~\cite{Matthews-Strogatz-90,Matthews-Mirollo-Strogatz-91}
could shed light on this problem. A promising approach for future
studies is a calculation of
finite-size Lyapunov exponents like in~\cite{Shibata98a,Cencini99}.

Another peculiar property of the system studied is its sensitivity
to the number of oscillators in the ensemble and appearance of periodic
``windows'' in dependence on this parameter. Such a sensitivity has not
been reported for other models demonstrating collective chaos. We
speculate that this property might be related to the mentioned above
existence of one positive Lyapunov exponent, what makes the collective
chaos in the ensemble less robust. This issue requires special attention
in the future work.

We believe, that the model proposed, although rather
artificial to be observed in natural oscillator ensembles, provides a
useful test system for analysis. On the other hand, 
our research opens a possibility of constructing realistic 
systems with collective chaotic phase dynamics based on ensembles of 
such individual elements  that show only regular dynamics.  
This might be feasible, e.g., on the basis of electronic devices, 
such as arrays of Josephson 
junctions
%contacts 
\cite{Wiesenfeld-Swift-95}, or with nonlinear optical systems, such as arrays of 
semiconductor lasers \cite{Glova-03}. Such systems 
are expected to generate robust 
chaos providing the power level much higher than that
characteristic for the 
individual elements. Systems of this kind may be of interest for 
applications requiring generation of chaotic signals, such as 
communication schemes \cite{Dmitriev-Panas-02,Koronovskii-Moskalenko-Hramov-09}, 
noise radar \cite{Lukin-01}, etc.

\begin{acknowledgments}
The research is supported, in part, by RFBR-DFG grant No 08-02-91963.
\end{acknowledgments}

%\bibliography{PhaseChaos_resub}
%

\end{document}